\documentclass{osa-article}

\journal{oe}

\usepackage{bm}
\usepackage{dsfont}
\usepackage{mathrsfs}
\usepackage{amsmath}     
\usepackage{amsfonts}  
\usepackage{amssymb}

\newcommand{\be}{\begin{equation}}
\newcommand{\ee}{\end{equation}}
\newcommand{\bea}{\begin{eqnarray}}
\newcommand{\eea}{\end{eqnarray}}
\newcommand{\aver}[1]{\langle #1 \rangle}
\newcommand{\ket}[1]{|#1\rangle}

\begin{document}

\title{Optical scheme for cryptographic commitments with physical unclonable keys}

\author{Georgios M. Nikolopoulos\authormark{1,2,*}}

\address{\authormark{1}Institute of Electronic Structure \& Laser, 
FORTH, P.O. Box 1385, GR-70013 Heraklion, Greece\\
\authormark{2}Institut f\"ur Angewandte Physik, Technische Universit\"at Darmstadt, D-64289 Darmstadt, Germany}

\email{\authormark{*}nikolg@iesl.forth.gr} 



\begin{abstract}
We investigate the possibility of using multiple-scattering optical 
media, as resources of randomness in cryptographic tasks pertaining to commitments and auctions. The proposed commitment protocol exploits 
standard wavefront-shaping and 
heterodyne-detection techniques, and can be implemented with 
current technology.  Its security is discussed in the framework of a tamper-resistant trusted setup. 
\end{abstract}


\section{Introduction}
Physical unclonable keys (PUKs) can be materialized by various types of optical multiple-scattering media. The scattering of coherent light from PUKs is a linear process involving a large number of optical modes, and leads to a speckle pattern that stems from the interference of many paths that lead to a particular mode at the output \cite{Goodman1}. 
Propagation of light in PUKs can be controlled by means of standard wavefront-shaping techniques, which allow for the control of hundreds, and even thousands of the supported transverse optical modes \cite{Vellekoop15,Mosk12,Poppoff11}.  

The internal disorder of a PUK on the one hand renders its cloning a formidable challenge, and on the other hand may 
serve as a source of randomness in the design of optical cryptographic protocols. So far, related studies have focused on the authentication of entities and messages
 \cite{Pappu02,Goorden14,Iesl16,NikDiaSciRep17,Nik18,Mes18}, the storage of cryptographic keys \cite{Horstmayer13},  document and package fingerprinting \cite{Buch05,Yeh12}, and key agreement \cite{Amit18}. In all of these studies, the proposed cryptographic protocols exploit, one way or another, the sensitivity of the speckle to various parameters of the setup, including the PUK and parameters of the light that is scattered from it.
 
Here, we investigate the possibility of using PUKs for cryptographic commitments, which are of particular importance 
in digital auctions, electronic voting, etc. Cryptographic commitment 
can be thought of as a binding promise, and is provided by the 
so-called commitment schemes \cite{Goldreich, Smart}. To describe the main stages of a commitment protocol, let us consider a realistic scenario pertaining to 
an auction, in which  all of the participants are located at the same place. During the {\em commit phase},  each bidder commits to a secret bid, and sends evidence for his commitment to the auctioneer. 
In the {\em reveal  phase},  the participants announce publicly their bids, and 
the auctioneer sells the item to the highest bidder, provided that the highest bid is consistent with the corresponding evidence that the auctioneer received in the commit phase. 
For an undisputed outcome of the auction, the commitment scheme adopted during the auction has to have three main properties. 
\begin{enumerate}
\item {\em Correctness}.  If the participants and the auctioneer behave honestly,   
and follow the protocol, then the auctioneer will learn the actual bids 
of the participants at the end of the reveal phase. 
\item {\em Hiding (Concealing)}.  The auctioneer  learns nothing about the bids, until the participants disclose them in the reveal phase.
\item {\em Binding}. A bidder cannot change his bid after the commit phase.
\end{enumerate}
To facilitate the following discussion, we will focus on the auctioneer (Alice) and one bidder (Bob).
The hiding (concealing) property ensures that  Alice cannot affect the outcome of the auction, because she does not know the bids until they are publicly announced. The binding property does not allow Bob 
to affect the final outcome by changing his bid. 

In conventional cryptography one cannot have a commitment scheme which is perfectly binding and perfectly hiding simultaneously \cite{Smart}. That is, at least one of the two properties can be ensured only under certain assumptions about the (computational) power of the receiver or the sender. Interestingly enough, the same is true in a quantum setting, where the secrets are encoded in the states of quantum systems \cite{BeyondQKD}. 
There are many approaches for the construction of numerical  commitment schemes, but one-way hash functions offer a rather simple and intuitive tool to this end. 
Let ${\cal F}$ be a publicly known one-way collision-resistant hash function.  This means that it is computationally easy to find $y={\cal F}(x)$ for a given $x$, but it is hard to find $x={\cal F}^{-1}(y)$ for given $y$. Moreover, 
 it is hard to find distinct $x_1$ and $x_2$ that yield the same result $y={\cal F}(x_1)={\cal F}(x_2)$. A commitment scheme can be built on this 
 function as follows. 
To commit to a secret bid $s$, Bob calculates ${\cal C}={\cal F}(s)$ and sends the commitment ${\cal C}$ to Alice.  In the reveal phase, Bob announces publicly his bid, and Alice can confirm that he is not cheating by calculating ${\cal F}(s)$ and by checking that the result matches ${\cal C}$ that Bob sent her in the 
commit stage. This very simple commitment protocol may look perfect, but it is not concealing when Alice can focus on a small number of  possible bids. To prevent this attack, one has to add additional randomness in the scheme, thereby making the choice of the secret more unpredictable.

Given that PUKs are inherently random, the question arises  whether one can exploit this randomness to the design of cryptographic  commitment schemes. Indeed, PUKs have been shown to behave as physical one-way functions, when one exploits the properties of the  random speckle obtained by scattering coherent light from them \cite{Pappu02, Goorden14,NikDiaSciRep17,Nik18,Horstmayer13}. In this work, we propose for the first time an optical commitment scheme that relies on PUKs, and  requires off-the-shelf optical components and standard techniques. Contrary to many  quantum commitment schemes, the proposed scheme is not limited to bits, and allows for the commitment to secrets of higher dimension.

In the following section we describe the main setup under consideration, and we  introduce the necessary formalism.  In Sec. \ref{Sec3} we present the commitment protocol, and discuss its security in the framework of a trusted tamper-resistant setup. A summary with concluding remarks is given in Sec. \ref{Sec4}. 

\begin{figure}
\centering\includegraphics[scale=0.4]{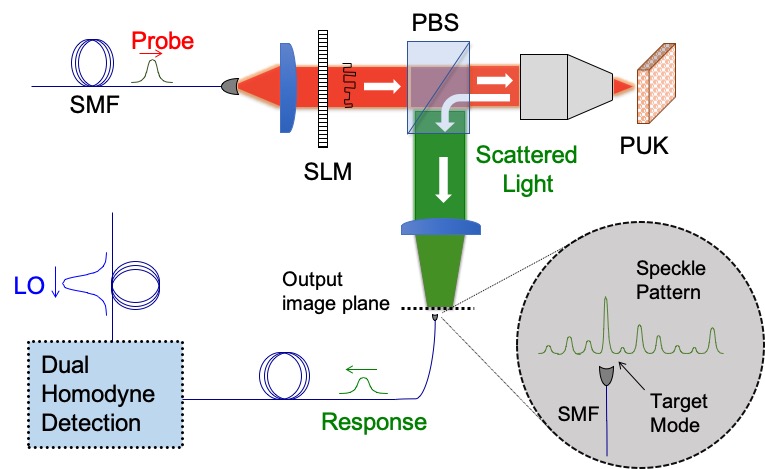}
\caption{
Schematic representation of the PUK analyzer considered in the proposed commitment scheme. The transverse spatial wavefront of the  probe  is shaped by means of  phase-only spatial light modulator (SLM). The shaped light is focused on the random PUK, and the scattered (reflected) light is collected by means of a polarizing beam splitter (PBS). The output 
field is imaged onto a plane, where a single-mode fiber (SMF) can be translated in a controlled manner, and the phase-mask of 
the SLM is optimized so that the speckle exhibits a single enhanced speckle grain, at the position of the SMF (target mode) \cite{Defienne14}. The overall imaging system is optimized so that the SMF collects light from a single speckle grain. 
A joint measurement of both quadratures of the collected light is performed by means of a  balanced dual-homodyne detection system. 
}
\label{fig1}
\end{figure}


\section{Setup and formalism}
\label{Sec2}
A schematic presentation of the setup under consideration is shown 
in Fig. \ref{fig1}.  Besides the detection system, the setup is analogous to the one used in connection with various possible applications of PUKs \cite{NikDiaSciRep17,Defienne14,Def16}. Incoming light in a coherent state $\ket{\alpha} := \ket{\sqrt{\mu}}$ is collimated and its wavefront is shaped by means of a 
phase-only spatial light modulator (SLM). The mean number of photons in the state is denoted by $\mu$, and without loss of generality, the corresponding phase is set to zero.  Using an objective, the  shaped light is focused on a PUK, with mean-free path $l$ and thickness $L$, and the scattered light (speckle) 
is collected by means of a polarizing beam splitter (which ensures collection of multiple-scattered light) and an objective. One of the speckle grains is coupled to a SMF (target mode), which leads to a standard dual-homodyne detection (DHD) setup, where the two conjugate quadrature  components of the mode of the SMF $(\hat{X}_s,\hat{Y}_s):=\hat{\bm Z}_s$, are analyzed simultaneously \cite{book3}. Introducing the bosonic annihilation operator for the target mode 
$\hat{b}_s$, one has 
$\hat{X}_s = (\hat{b}_s+\hat{b}_s^\dag)/\sqrt{2}$ and $\hat{Y}_s = (\hat{b}_s-\hat{b}_s^\dag)/\sqrt{2}$. In the linear diffussive regime, the scattering can be described in the framework of input-output  formalism obtaining \cite{NikDiaSciRep17,Goodman1, Lodahl_prl05,Lodahl_oe06}
\bea
\aver{\hat{b}_s} =\sum_{j=1}^N r_{s,j} e^{{\rm i}\phi_j}\aver{\hat{a}_j},
\eea
where $N$ is number of independent transverse spatial modes at the input, and  $\hat{a}_j$ denotes  the  annihilation operator for the $j$-th input mode.  The elements $\{r_{s,j}\}$ pertain to the $s$-th row of the reflection matrix of the PUK, and they are statistically independent complex Gaussian random variables. 
The phases $\Phi:=\{\phi_j\}$ constitute the phase mask of the SLM. 
Assuming a uniform illumination of the SLM we have 
$\aver{\hat{a}_j} = \alpha\sqrt{\tau/N}$, 
where $\tau<1$ accounts for losses, while the coherence of the input state is preserved throughout its propagation in the system. For fixed  randomly chosen phase mask, the quantum mechanical expectation value $\aver{\hat{\bm Z}_s}$ is a bivariate normal random variable that  varies with the realization of disorder and satisfies 
\cite{NikDiaSciRep17,Goodman1, Lodahl_prl05,Lodahl_oe06}
\bea
\overline{\aver{{\bm Z}_s}} = (0,0), \quad {\rm Var}(\hat{\bm Z}_s) = 
2\mu\frac{\tau}{N} \left (1-\frac{l}L\right):= 2{\mu}{\cal V},
\label{Zstats}
\eea
where  ${\rm Var}(\hat{\bm Z}_s):= \overline{(\aver{\hat{X}_s})^2}+\overline{(\aver{\hat{Y}_s})^2}$, and the overline denotes ensemble average. As a result, if one was able to obtain the exact value of $\aver{\hat{\bm Z}_s}$ for  different realizations of disorder  then, with high probability, the  recorded values  for the different realizations would lie within or very close to a circular area  of radius 
\be
\rho:=4\sqrt{\mu{\cal V}}
\label{rho:eq}
\ee
around the origin $(0,0)$. 
DHD of the scattered field at the output is equivalent to sampling from the  normal distribution 
${\cal N}(\aver{\hat{\bm Z}_s}, {\bm \sigma})$ with covariance matrix 
${\bm \sigma} = {\bm 1}/\sqrt{\eta}$, where ${\bm 1}$ is the identity $2\times 2$ matrix, and $0.5\leq \eta<1$ is the detection efficiency.

\begin{figure}
\centering\includegraphics[scale=0.33]{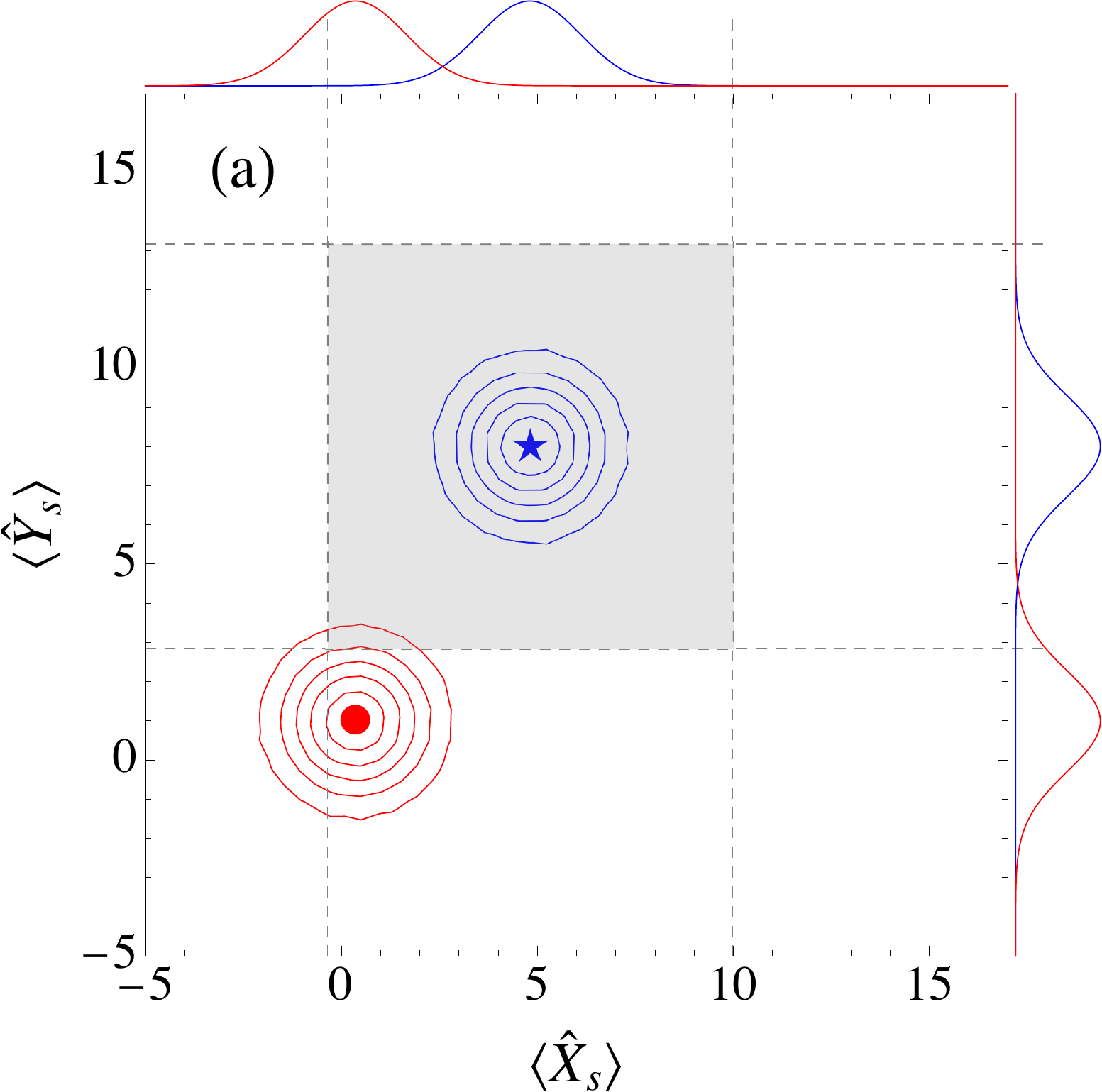}
\includegraphics[scale=0.33]{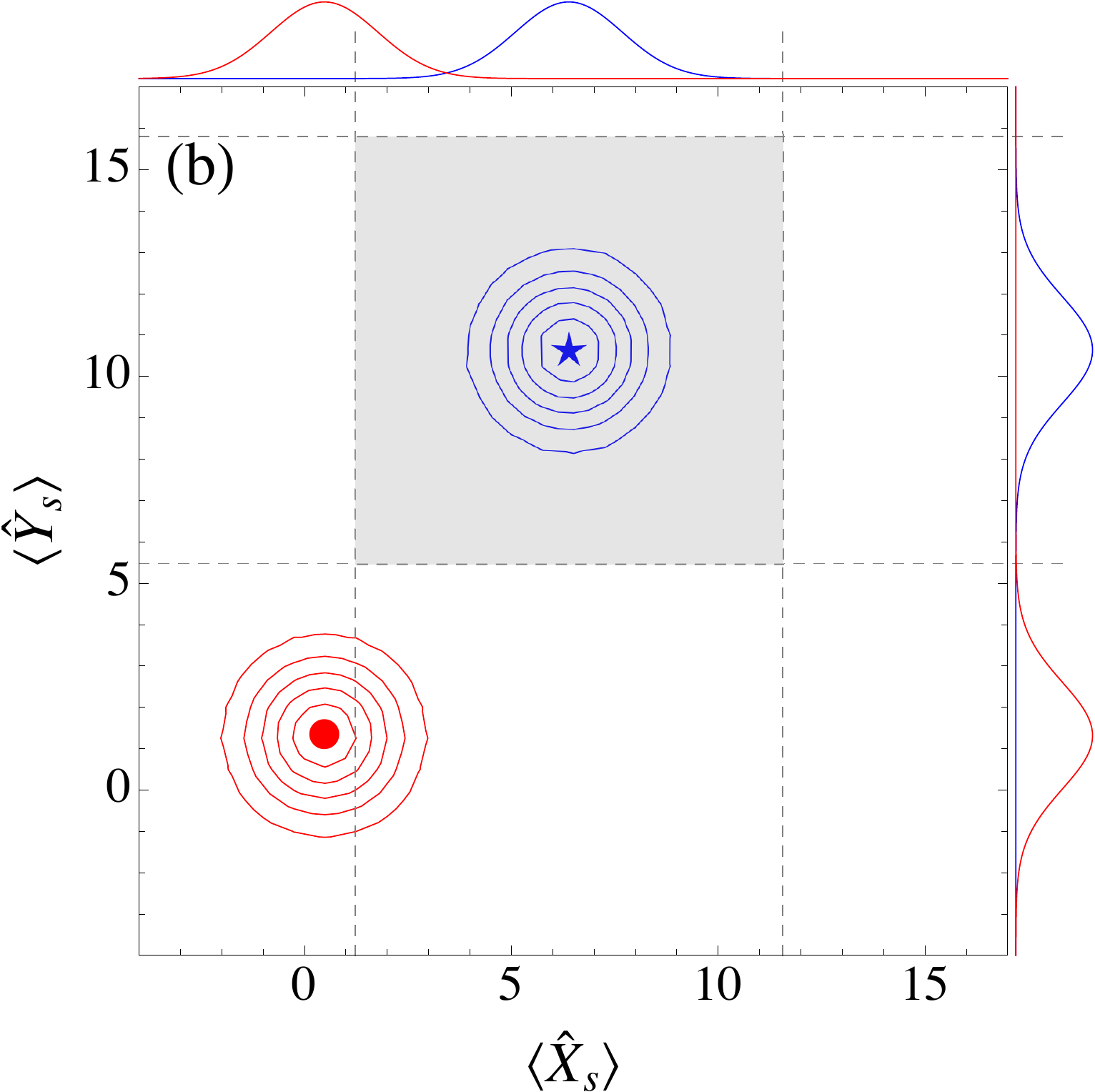}
\caption{
Contour plot of the typical response of a PUK for optimized (blue) and non-optimized (red) SLM, for two different values of the mean number of photons: 
(a) $\mu=1500$, and (b) $\mu=2650$. The dashed vertical and horizontal lines show the overlap of the corresponding marginal distributions for the quadratures, with the acceptance region ${\mathscr A}$ (gray rectangle). Parameters:  $N=625$, $w = 8/\sqrt{\eta}$, $l/L=0.2$, $\tau = 0.05$, $\eta=0.6$.}
\label{fig2}
\end{figure}

By applying standard techniques, for a fixed PUK one can optimize the phase-mask of the SLM so that the intensity of the scattered field in the SMF is maximized \cite{Vellekoop15,Mosk12,Poppoff11}. In this case, we have for the optimized response 
$\aver{\hat{\bm Z}_s}_{\rm o}:=(\aver{\hat{X}_s}_{\rm o},\aver{\hat{Y}_s}_{\rm o})$ 
the constraint 
\bea
(\aver{\hat{X}_s}_{\rm o})^2+(\aver{\hat{Y}_s}_{\rm o})^2 = 2 {\cal E} \mu {\cal V} := 2\rho_{\rm o}^2,
\label{rho0:eq}
\eea
where the enhancement factor ${\cal E}$  quantifies the achievable control  on the scattering of the light in the particular setup.   Typical enhancements that have been reported in the literature range from about $10$ to more than $10^3$, depending mainly on the number of controlled modes, the type of light modulation, the intensity distribution over the input modes, and the stability of the scattering medium \cite{Vellekoop15,Yilmaz13,Anderson14}. 
In the case of optimized SLM, the sample is obtained from a normal distribution ${\cal N}(\aver{\hat{\bm Z}_s}_{\rm o}, {\bm \sigma})$, and constraint (\ref{rho0:eq}) implies that either $|\aver{\hat{X}_s}_{\rm o}|\geq \rho_{\rm o}$ or $|\aver{\hat{Y}_s}_{\rm o}|\geq \rho_{\rm o}$. 

The  protocol that is proposed in the following section relies on the discrimination between the optimized and the non-optimized case by means of the DHD, which is equivalent to distinguishing between 
the   distributions ${\cal N}(\aver{\hat{\bm Z}_s}_{\rm o}, {\bm \sigma})$ and ${\cal N}(\aver{\hat{\bm Z}_s}, {\bm \sigma})$, through sampling. 
When sampling from either of the distributions, the outcome of a single DHD is 
a bivariate random variable ${\bm z}=(x,y)$, and decision making can be based e.g., on a rectangular acceptance region 
\begin{equation*}
  {\mathscr A}(\widetilde{\bm Z}_s^{(\rm o)},w):= \left\{
    \begin{array}{r}
      \widetilde{X}_s^{(\rm o)} - w/2\leq  x\leq \widetilde{X}_s^{(\rm o)} + w/2\\
      \widetilde{Y}_s^{(\rm o)} - w/2 \leq y 
 \leq \widetilde{Y}_s^{(\rm o)} + w/2
    \end{array} \right \},
\end{equation*}
which is centered at a reliable estimate of $\aver{\hat{\bm Z}_s}_{\rm o}$, say $\widetilde{\bm Z}_s^{(\rm o)}:=( \widetilde{X}_s^{(\rm o)},\widetilde{Y}_s^{(\rm o)})\simeq\aver{\hat{\bm Z}_s}_{\rm o}$, and 
has width $w$ (see Fig. \ref{fig2}). If the outcome of a single DHD 
falls in the region ${\mathscr A}$ we conclude that the sample has been obtained from 
${\cal N}(\aver{\hat{\bm Z}_s}_{\rm o}, {\bm \sigma})$, and in the opposite case from 
${\cal N}(\aver{\hat{\bm Z}_s}, {\bm \sigma})$. Assuming that 
$|\aver{\hat{\bm Z}_s}_{\rm o}- \widetilde{\bm Z}_s^{(\rm o)}|\ll 1/\sqrt{\eta}$, the sample size required for reliable decision making depends on the distance between $\aver{\hat{\bm Z}_s}_{\rm o}$ and $\aver{\hat{\bm Z}_s}$, relative to the width $w$  and the detection efficiency $\eta$.  More precisely, the 
probability for an outcome to fall in the acceptance region when sampling from ${\cal N}(\aver{\hat{\bm Z}_s}, {\bm \sigma})$, is given by the integral of  ${\cal N}(\aver{\hat{\bm Z}_s}, {\bm \sigma})$ over $ {\mathscr A}$, which yields 
\bea
P_{\rm in}&=& \frac{1}{4}\left [  
{\rm Erf}\left (
\frac{\tilde{w}}{2\sqrt{2}}+
\frac{( \widetilde{X}_s^{\rm (o)} - \aver{\hat{X}_{s}})\sqrt{\eta}}{\sqrt{2}} 
\right )+{\rm Erf}\left (
\frac{\tilde{w}}{2\sqrt{2}}-
\frac{( \widetilde{X}_s^{\rm (o)} - \aver{\hat{X}_{s}})\sqrt{\eta}}{\sqrt{2}} 
\right )
\right ]\nonumber\\
&&\times 
\left [  
{\rm Erf}\left (
\frac{\tilde{w}}{2\sqrt{2}}+
\frac{( \widetilde{Y}_s^{\rm (o)} - \aver{\hat{Y}_{s}})\sqrt{\eta}}{\sqrt{2}} 
\right )+{\rm Erf}\left (
\frac{\tilde{w}}{2\sqrt{2}}-
\frac{( \widetilde{Y}_s^{\rm (o)} - \aver{\hat{Y}_{s}})\sqrt{\eta}}{\sqrt{2}} 
\right )
\right ]. 
\label{pinGeneral:eq}
\eea
where $\tilde{w}:=w\sqrt{\eta}$.
The distance between the two distributions is determined by the difference $\rho_{\rm o}-\rho$, which increases with the mean number of photons in the probe (see also Fig. \ref{fig2}). 
More precisely, using Eqs. (\ref{rho:eq}) and (\ref{rho0:eq}) one readily obtains that  for a chosen $\Omega>0$ we have 
\bea
\Delta:=(\rho_{\rm o}-\rho)\sqrt{\eta} \geq \Omega,
\label{Delta:eq}
\eea
for 
\bea
\mu  \geq \frac{\Omega^2}{\eta {\cal V}(\sqrt{{\cal E}}-4)^2}.
\label{critical_mu:eq}
\eea
Throughout this work, the width of the acceptance region, the mean number of photons, the detection efficiency, and various other parameters of the setup, are  considered to be publicly known. 

\begin{figure}
\centering\includegraphics[scale=0.33]{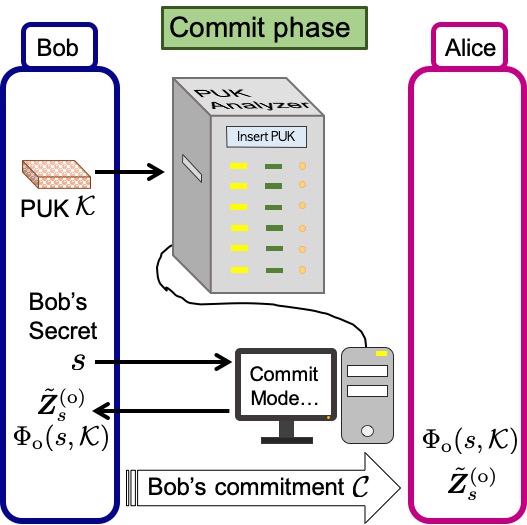}
\hspace*{0.3cm}
\includegraphics[scale=0.33]{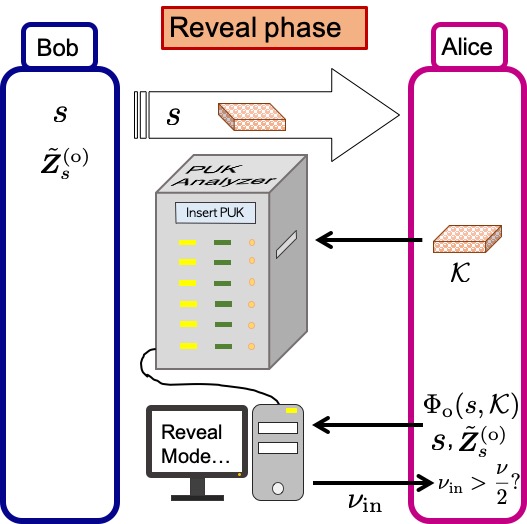}
\caption{
Schematic representation of the commit and the reveal phases in the commitment scheme under consideration. }
\label{fig3}
\end{figure}


\section{Protocol} 
\label{Sec3}
Consider  a box which contains random independently prepared PUKs, and the users have access to it during the protocol only. We assume that there are $n$ different accessible target modes in the setup, and they are labeled by an integer, which takes values in the set ${\mathbb S} = \{0, 1, \ldots, n-1\}$. 
The secret bid of Bob $s$ also takes values in ${\mathbb S}$, and identifies uniquely a target mode.  
The commitment protocol is summarized in Fig. \ref{fig3} and proceeds as follows.
\begin{itemize}
\item {\em Commit phase.}   Bob chooses at random a PUK ${\cal K}$, and inserts it to the  analyzer, together with his secret $s$. The analyzer optimizes the phase-mask of the SLM so that the intensity of the  scattered light is maximized at mode $s$. Subsequently, the analyzer interrogates the PUK with many probes in state $\ket{\alpha}$, and a  reliable estimate of $\aver{\hat{\bm Z}_s}_{\rm o}$ (to be denoted by $\widetilde{\bm Z}_s^{\rm (o)}$) is obtained by means of DHD of the scattered light in the SMF for each probe. Bob's commitment ${\cal C}:=\{\Phi_{\rm o}(s,{{\cal K}}), \widetilde{\bm Z}_s^{({\rm o})}\}$  is given to Alice, where 
$\Phi_{\rm o}(s,{{\cal K}})$ denotes the optimal phase mask for the given PUK-target pair $({\cal K},s)$.
\item {\em Reveal phase.} Bob discloses his bid $s$, and gives his PUK to Alice. To confirm Bob's value, Alice uses the same analyzer as in the commit phase. She sets the phase mask of the SLM to $\Phi_{\rm o}(s,{{\cal K}})$, and the target mode to $s$. Then she interrogates the PUK with $\nu\geq 1$ coherent pulses 
for some odd $\nu$, each one in  state $\ket{\alpha}$,  and applies DHD on the scattered light.  She accepts Bob's bid if the majority of the outcomes fall within the rectangular area ${\mathscr A}(\tilde{\bm Z}_s^{(\rm o)}, w)$, and rejects it otherwise. 
\end{itemize}

Having presented the protocol, in the following subsections we discuss its correctness, as well as its security with respect to the binding and the hiding properties. 


\subsection{Correctness}
When both users behave honestly, Alice should be able to confirm  Bob's true bid with high confidence at the reveal phase, with finite resources. The probability for Alice to accept Bob's true bid in the reveal phase is given by 
\bea
P_{\rm accept}^{(\nu)}:= \sum_{j=0}^{\lfloor \nu/2\rfloor} 
\binom{\nu}{j}[1-p_{\rm in}^{({\rm o})}]^{j}[p_{\rm in}^{({\rm o})}]^{\nu-j}, 
\label{Paccept1:eq}
\eea
where $p_{\rm in}^{({\rm o})}$ is the probability for a random outcome of the DHD to fall within the acceptance region ${\mathscr A}(\tilde{\bm Z}_s^{\rm (o)}, w)$, when sampling from ${\cal N}(\aver{\hat{\bm Z}_s}_{\rm o},{\bm \sigma})$. To ensure reliable decision making, one needs first a rather precise estimate of $\aver{\hat{\bm Z}_s}_{\rm o}$ in the commit phase,  which can be obtained through sampling. By interrogating the PUK with 
$M$ coherent states $\ket{\alpha}$, one can approximate  $\aver{\hat{\bm Z}_s}_{\rm o}$ 
by the mean value of the recorded outcomes. The sample mean  $\widetilde{\bm Z}_s^{(\rm o)}$ is a random variable which, according to the central-limit theorem, follows  a normal distribution centered at $\aver{\hat{\bm Z}_s}_{{\rm o}}$, and  with covariance matrix ${\bm \sigma}/\sqrt{M}$.  
For a moderate number of experiments, say $M\approx 10^3$, one can ensure that the error for either of the two quadratures is much smaller than $1/\sqrt{\eta}$, with high confidence $(\sim 99.99\%)$, and the acceptance region ${\mathscr A}$ will be centered pretty close to $\aver{\hat{\bm Z}_s}_{\rm o}$. The probability $p_{\rm in}^{({\rm o})}$ is given by Eq. (\ref{pinGeneral:eq}), when $\aver{\hat{\bm Z}_s}$ 
is replaced by $\aver{\hat{\bm Z}_s}_{\rm o}\simeq \widetilde{\bm Z}_s^{(\rm o)}$, thereby obtaining  
\bea
p_{\rm in}^{({\rm o})} \simeq \left[{\rm Erf}\left (\frac{\tilde{w}}{2\sqrt{2}} \right )\right ]^2,
\label{Paccept2:eq}
\eea
which depends solely on the width of the acceptance region relative to the detection efficiency, and it increases monotonically with $\tilde{w}$. 
Any value of $\tilde{w} \geq 7$ ensures $p_{\rm in}^{({\rm o})} \simeq 1$, and for the 
sake of concreteness throughout our simulations we  considered $\tilde{w} =  8$, 
so that  $1-p_{\rm in}^{({\rm o})}\simeq 10^{-4}$ and 
$P_{\rm accept}^{(\nu)}\simeq 1$ for any  $\nu\geq 1$.


\subsection{Concealing}
At the end of the commit phase, Alice receives Bob's commitment ${\cal C}$. 
In order for Alice to deduce Bob's secret, she has to deduce the target mode from it. However, this is not possible without access to Bob's PUK. The optimal phase mask carries information about the phases of the elements in a specific row of the reflection (or  transmission) matrix, and the row is determined by  Bob's secret. For a strongly scattering PUK, the elements of the matrix  are statistically independent complex Gaussian random variables, and thus do not convey any information about their position in the matrix (i.e., about the label of the row).  Indeed, any permutation of the rows of the reflection matrix results in a valid reflection matrix, and thus to a valid PUK-target pair. So, every value in ${\mathbb S}$ is equally likely to be Bob's bid, given what Alice has seen.  If Alice has unlimited power of  computation is not expected to help her either. Hence, the scheme is  concealing, in the sense that the probability for Alice to deduce Bob's secret is not better than the probability of random guessing  $p_{\rm rg} = 1/n$.  

\begin{figure}
\centering\includegraphics[scale=0.5]{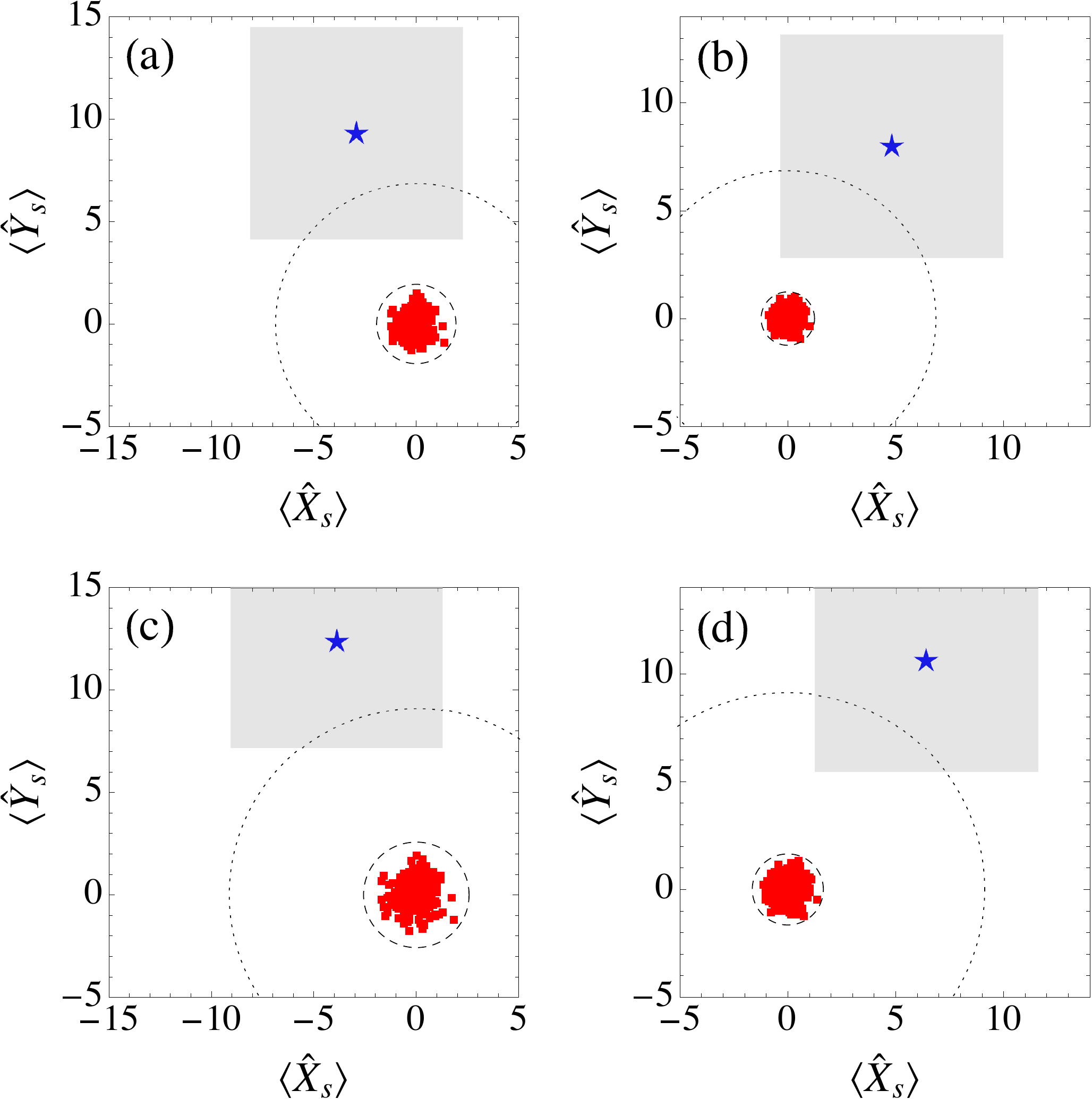}
\caption{
Phase-space representation of the response of a PUK for optimized 
and non-optimized SLM. The blue star shows the response $\aver{\hat{\bm Z}_s}_{\rm o}$ for the target mode $s$,  according to which the phase mask of the SLM has been optimized. 
The red disks show the response of the PUK for all of the other output modes $s^\prime\neq s$, while keeping the phase mask to its optimal configuration for $s$.   
(a) $N=256$, $\mu=1500$; (b) $N=625$, $\mu=1500$;  (c) $N=256$, $\mu=2650$, (d) $N=625$, $\mu=2650$. The dashed curves show circular areas of radius $\rho_{\rm o}$ and 
$\rho$, with $\rho_{\rm o}>\rho$. The gray area shows the acceptance region ${\mathscr A}(\aver{\hat{\bm Z}_{s}}_{\rm o}, 8/\sqrt{\eta})$. Other  parameters:  $l/L=0.2$, $\tau = 0.05$, $\eta=0.6$. }
\label{fig4}
\end{figure}


\subsection{Binding}
Recall now that Bob's commitment to bid $s$ involves the optimal SLM mask  $\Phi_{\rm o}(s,{\cal K})$, which depends on the target mode (i.e., the bid) and on the used PUK. Bob cannot alter the phase mask after he gives it to Alice. Hence, in order to cheat successfully, he has to find  beforehand (during the commit phase), an additional target mode $s^\prime\neq s$ for which the expected response of the given PUK under the  phase mask $\Phi_{\rm o}(s,{\cal K})$ lies sufficiently close to the acceptance region ${\mathscr A}(\tilde{\bm Z}_{s}^{(\rm o)}, 8/\sqrt{\eta})$, so that to result in a non-negligible probability for Alice to accept $s^\prime$. Our simulations show that  for a fixed PUK, the optimal phase mask $\Phi_{\rm o}(s,{\cal K})$ is very sensitive to the target mode $s$ for which the optimization has been performed. Hence, $\Phi_{\rm o}(s,{\cal K})$ acts as a totally random phase mask for all of the output modes $s^\prime\neq s$, and the corresponding responses lie close to the origin (0,0), within a circular area of radius $\rho$ (see Fig. \ref{fig4}). 
The probability of successful cheating $P_{\rm cheat}^{(\nu)}$, depends on  the position of the center of the normal distribution ${\cal N}(\aver{\hat{\bm Z}_{s^\prime}}, {\bm \sigma})$, relative to the acceptance region ${\mathscr A}(\tilde{\bm Z}_s^{\rm (o)},w)$. In particular, $P_{\rm cheat}^{(\nu)}$ is given by
\begin{subequations}
\label{UpperBound}
\bea
P_{\rm cheat}^{(\nu)}:= \sum_{j=0}^{\lfloor \nu/2\rfloor} 
\binom{\nu}{j}(1-p_{\rm in})^{j}(p_{\rm in})^{\nu-j}
\label{PCheat:eq}
\eea
where $p_{\rm in}$ is the probability for Bob's false bid $s^\prime$ to result in an outcome within the acceptance region in a single experiment. It is given by Eq. (\ref{pinGeneral:eq}) after replacing $\aver{\hat{\bm Z}_{s}}$ 
by $\aver{\hat{\bm Z}_{s^\prime}}$.

\begin{figure}
\centering\includegraphics[scale=0.5]{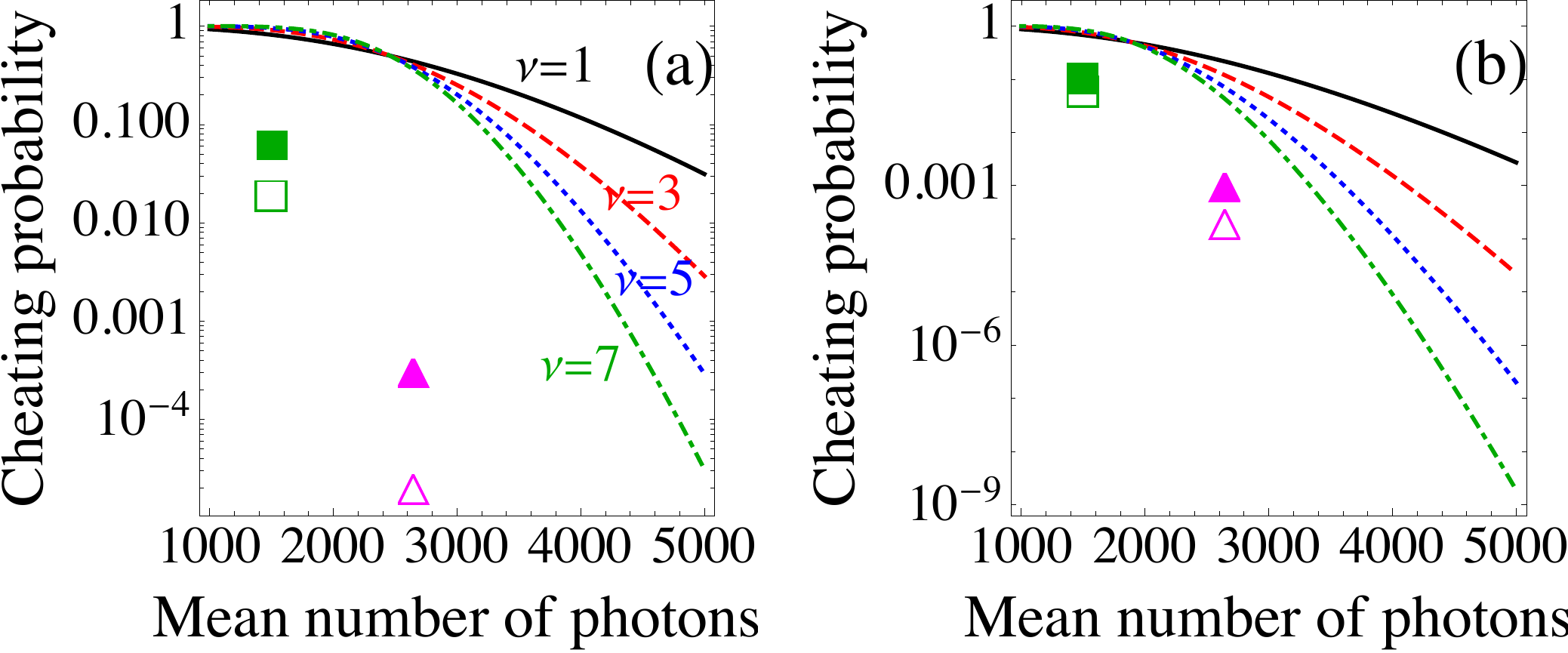}
\caption{
Probability of successful cheating as a function of the mean number of photons in the pulse for $N=256$ (a) and $N=625$ (b). The curves show the theoretical upper bound on the probability of successful cheating for different values 
of $\nu$.  
The symbols show the maximum probabilities that have been extracted 
from the numerical data of Fig. \ref{fig4} (empty symbols) and Fig. \ref{fig6} (filled symbols) in the case of $\nu=1$, and they are  always below the theoretically expected bound (solid black curve). Squares (empty and filled) refer to $\mu=1500$ and the triangles to $\mu=2650$. Other  parameters as in Fig. \ref{fig4}.}
\label{fig5}
\end{figure}

The cheating probability $P_{\rm cheat}^{(\nu)}$ increases monotonically with 
$p_{\rm in}$, and in order to obtain an upper bound on it, we can maximize $p_{\rm in}$, with respect to $\widetilde{\bm Z}_s^{\rm (o)} = [\rho_{\rm o}\cos(\phi),\rho_{\rm o}\sin(\phi) ]$ and $\aver{\hat{\bm Z}_{s^\prime}} = [\rho\cos(\psi),\rho\sin(\psi) ]$. In these expressions we have taken into account that, with high probability, the response of the PUK for all $s^\prime\neq s$, lies within distance $\rho$ from the origin, whereas the response of the actual target mode $s$ is found at a distance $\rho_{\rm o}$ from the origin. One can readily confirm that $p_{\rm in}$ is maximized when $\widetilde{\bm Z}_s^{\rm (o)} $ and $ \aver{\hat{\bm Z}_{s^\prime}}$ 
are aligned, and $\phi = \psi = l\pi/4$, for $l=1,3,5,7$. The maximum value is 
\bea
p_{\rm in}^{({\rm max})} =\frac{1}{4}\left [  
{\rm Erf}\left (
\frac{\tilde{w}}{2\sqrt{2}}+
\frac{\Delta}{2} 
\right )+{\rm Erf}\left (
\frac{\tilde{w}}{2\sqrt{2}}-
\frac{\Delta}{2} 
\right )
\right ]^2,
\label{pinCheat_max:eq}
\eea
\end{subequations}
which when inserted in Eq. (\ref{PCheat:eq}),  yields an upper bound on $P_{\rm cheat}^{(\nu)}$. In order for the bound to be meaningful, one has to choose $\tilde{w}$ and $\Delta$ such that $p_{\rm in}^{({\rm max})}<0.5$. This is always possible, because for a given detection efficiency, $w$ is chosen at will and $\Delta$ is determined by the mean number of photons in the probe. 
As shown in Fig. \ref{fig5},  for given $\nu$ the upper bound can become arbitrarily small (while keeping $P_{\rm accept}^{(\nu)}\simeq 1$), by choosing sufficiently large mean number of photons in the probe.  It is also worth noting here that  the protocol becomes deterministic for sufficiently large values of $\mu$,  in the sense that $P_{\rm cheat}^{(\nu=1)}\ll 1$. For instance, as depicted in Fig. \ref{fig5}, in the case of $\mu=2650$ we have $P_{\rm cheat}^{(\nu=1)} <  10^{-3}$ for both $N=256$ and $N=625$ (see empty and filled triangles).

\begin{figure}[t!]
\centering\includegraphics[scale=0.5]{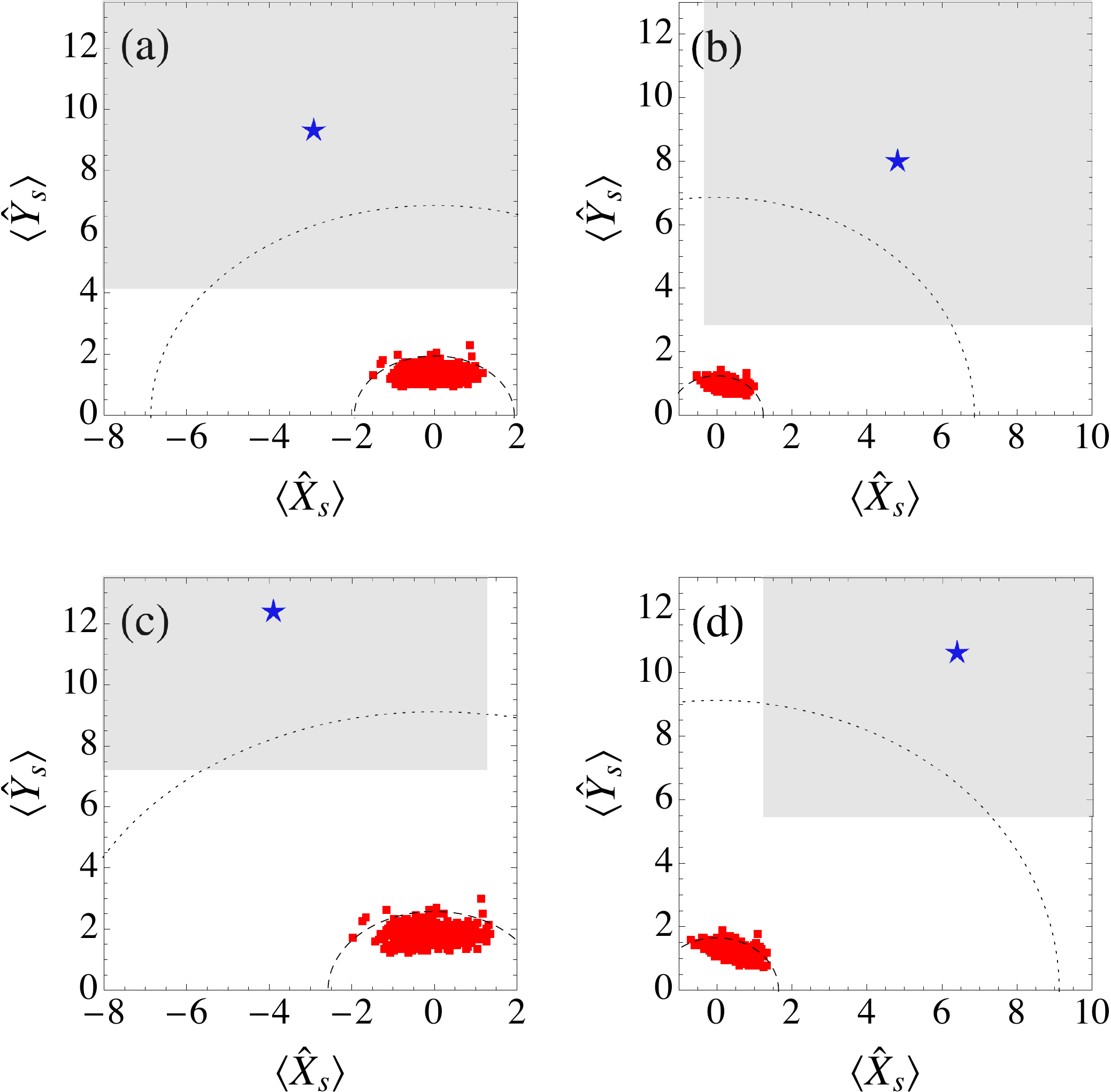}
\caption{
Phase-space representation of the response of the PUK for optimized 
and non-optimized SLM. The blue star shows the response for the reference 
PUK-target pair $({\cal K},s)$, according to which the SLM has been optimized. The red disks show the 
responses of PUK-target pairs $({\cal K}^\prime,s^\prime)$ that maximize the probability of acceptance, when the phase-mask of SLM is set to its optimal configuration for $({\cal K},s)$. The data have been obtained from simulations on 500 random keys ${\cal K}^\prime$, and all of the  possible output modes $s^\prime \neq s$. Other parameters as in Fig. \ref{fig4}. }
\label{fig6}
\end{figure}

The question arises whether Bob can improve on his probability of successful cheating, if he is not limited to a single PUK, and moreover he has unlimited access to a tamper-resistant analyzer.  
For instance, in this case he may try  to find a common optimal phase mask for two distinct PUK-target pairs, say $({\cal K}, s)$  and $({\cal K}^\prime, s^\prime)$ for some $s^\prime\neq s$. To investigate this cheating strategy, we optimized the SLM with respect to a reference PUK-target pair $({\cal K}, s)$, thereby obtaining the optimal phase mask $\Phi_{\rm o}(s,{\cal K})$, and the response 
$\aver{\hat{\bm Z}_{s}}_{\rm o}$. Subsequently, we generated 500 random PUKs 
(${\cal K}^\prime$), and for each one of them we calculated the response for all of the possible output modes $s^\prime\neq s$, and with the SLM configuration fixed to $\Phi_{\rm o}(s,{\cal K})$. For each PUK-target pair $({\cal K}^\prime, s^\prime)$ we recorded the response that resulted to the largest probability for an outcome to fall in the acceptance region, and thus to the largest probability of successful cheating.  As shown in Fig. \ref{fig6}, the best responses tend to align with the response of the reference pair $({\cal K}, s)$, but they are always concentrated at the periphery of the circle with center $(0,0)$ and radius $\rho$. As a result, there is only a small improvement on the probability of successful cheating relative to the aforementioned scenario, where Bob is limited to choose only one PUK during the commit phase (compare empty and filled symbols of the same type in Fig. \ref{fig5}).  
Moreover, the estimated probability for successful cheating is always below the upper bound given by Eqs. (\ref{UpperBound}).


\section{Discussion}
\label{Sec4}
We have proposed a commitment protocol that relies on PUKs, and can be implemented with current technology and off-the-shelf optical components. In particular, using standard techniques the wavefront of the incoming light is optimized so that to maximize the intensity of the scattered light at a particular target mode at the output, which is determined by the secret message (bid). Choosing sufficiently large the mean number of photons in the probe, the scheme becomes almost deterministic, in the sense that reliable decision making in the  reveal phase  can be conducted with the scattering of a single pulse from the PUK and DHD on the scattered light. The operation of the protocol to smaller mean number of photons requires a moderate number of measurements.  If necessary,  it is straightforward to modify the proposed protocol so that decision making relies on standard statistical techniques, such as hypothesis tests and confidence intervals (e.g., see  \cite{NikDiaSciRep17}).  In this case one may need a larger 
(not prohibitive) number of measurements, 
but there may be more freedom with respect to the choice of various parameters such as the width of the acceptance region, and the mean number of photons. 

The security of the protocol has been discussed under the assumption of tamper-resistant and trusted PUK analyzer. In particular, we assumed that  the interface  between the  analyzer and the user prohibits other 
actions on a given PUK, besides the optimization of the SLM with respect to the intensity of the scattered light at a single target mode. 
The protocol is  hiding, due to the independence of the random elements in the scattering matrix of the PUK,  while the binding property stems from the strong dependence of the optimal wavefront on the internal disorder of the PUK, and on the target mode.  

The security of the protocol under more general scenarios is a subject of future work, which goes beyond the scope of the present work. 
For instance, here we assumed that during the commit phase the interface  between the PUK analyzer and Bob is such that it accepts as an input only one integer $s\in {\mathbb S}$, and the PUK ${\cal K}$, while it outputs only $\{\Phi_{\rm o}(s, {\cal K}); \tilde{\bm Z}_s^{\rm (o)}\}$.  Although this is a very reasonable assumption, which can be readily satisfied through the software that provides the interface between the user and the analyzer, it is worth asking what happens if one relaxes this constraint, thereby giving Bob unrestricted access to the setup during the commit phase. In this case Bob may, for instance, optimize the SLM with respect to two target modes simultaneously, so that the overall setup operates as a beam splitter \cite{Huisman:14}.   Subsequently, during the reveal phase, he can decide on which of the two choices will be used. The question arises whether Bob can cheat successfully in this scenario, and if yes, whether the protocol can be modified so that to become secure again. 
Analogous questions can be considered when Alice has unrestricted access 
 to the analyzer. In this case, the encryption of the optimal phase mask during
 the commit phase, may improve the security of the protocol with respect to 
 its hiding property. 
Moreover, in this work we  assumed that identical  PUK analyzers  are used in the two  phases of the protocol. The security of the protocol in the case where the PUK analyzers are not precisely the same, remains to be investigated. In this case, inevitable deviations between the analyzers may, in principle, be exploited by a dishonest user. 
We believe that for sufficiently small deviations the parameters 
in the protocol, and in particular the width of the acceptance region, can be adjusted so that the security is not compromised.  

As mentioned above, the proposed protocol relies on a trusted authority, which fabricates the PUKs used in the protocol, and  certifies their randomness. First of all it is worth emphasizing that trusted authorities and centers play a pivotal role in various widely used cryptographic protocols \cite{handbook},  in the extension of quantum key-distribution links to intercontinental distances (e.g.,  earth-satellite links), as well as in the development of quantum pubic-key cryptography \cite{Nik2007}. Hence, the existence of a trusted authority does not in any case make our protocol less practical or less secure than many other quantum or conventional cryptographic protcools. The remaining question is how the trusted authority can ensure and verify the randomness of an optical PUK. This question is pertinent to any PUK-based cryptographic scheme, and to the best of our  knowledge, it has not been addressed in the literature so far. In most of the optical  PUKs that have been discussed in the literature so far, the internal  randomness is due to uncontrolled byproducts of the fabrication process and defects that are introduced intentionally during the fabrication of the PUK. In either case, this kind of randomness cannot be considered true, in the sense that it does not originate from a quantum process \cite{QRNG2017}. In fact, for certain  fabrication techniques particular care has to be taken so that to avoid systematic patterns in the PUKs (e.g., see  \cite{Pappu02}).  
Quantum physics is fundamentally random, and quantum systems are perfect sources of entropy. Hence, one way to ensure the randomness of the PUK is to combine standard laser-writing techniques, with certified quantum random-number generators (QRNGs), which are currently available on the market \cite{QRNG2017}. In particular, laser-writing methods allow for the fabrication of defects (modifications of the refractive index) on the surface or in the bulk of a glass, with rather precise positioning (up to few nanometers) and submicron-sized features (e.g., see \cite{Iesl16} and references therein). The position and the shape of the defects can be controlled by adjusting the parameters of the laser (e.g., position, intensity, illumination time, etc), and random variations of these parameters during the writing are essentially  imprinted onto the glass as random defects. Hence, by controlling the random variations of these parameters through  independent QRNGs, instead of the pseudorandom  number generators of a computer, one can ensure that the randomness of the resulting structure is the fingerprint of true randomness, which has originated from the QRNGs, and it has   been imprinted onto the glass by means of the laser. 

As far as the verification of the randomness is concerned, one may look at the correlations of the speckles corresponding to different positions of incidence. To this end, there is no need for wavefront shaping, and classical light can be used. Ideally, these speckles should be totally uncorrelated, and the deviation from the ideal scenario can be quantified by the correlation factor. Alternatively, one may also study the statistics of the light at various modes at the output, so that to confirm Eqs. (\ref{Zstats}), as well as the correlation between different modes \cite{Lodahl_prl05}. Such studies are expected to enable the quantification of possible deviations from the diffusive-limit and the Gaussian statistics model \cite{NikDiaSciRep17}, which are at the core of the present protocol. Finally, if necessary, one can also measure the scattering matrix of a given PUK by 
means of standard techniques (e.g. see \cite{Poppoff11} and references therein), which allows for the numerical analysis of the randomness. 

In closing, it is worth emphasizing that the proposed optical scheme is fundamentally different from its quantum counterparts that have been proposed in the literature so far \cite{BeyondQKD}. In analogy to conventional commitment schemes, unconditionally secure quantum bit commitment is impossible, but secure protocols can be designed in the framework of certain assumptions, such as the noisy-storage assumption \cite{Ng2012},  and the presence of four trusted agents who participate both in the commit and in the reveal phase \cite{Kent2012,Liu2014,Lunghi2013}. By contrast, the security of the present scheme relies on the  randomness of the PUK, as well as on the existence of a trusted PUK analyzer. Moreover, the present scheme does not require sources of entangled photon pairs or sources of single photons.  As a result, it is 
more efficient and practical than various quantum bit-commitment schemes that have been realized experimentally \cite{Ng2012,Liu2014,Lunghi2013}. The main advantage of the quantum bit commitment schemes in Refs. \cite{Ng2012,Liu2014,Lunghi2013} is that they can operate for distant parties, whereas throughout this work we have focused on a specific realistic  scenario, where both parties are located at the same place, and have access to the PUK analyzer. 
However, the extension of the present protocol to distant parties should not be considered impossible, and remains a very important open question that deserves further investigation. The techniques of Refs. \cite{Amit18,Def16,Leed18}, 
and the use of multimode fibers, may be particularly useful in this direction.

\section*{Funding}
Deutsche Forschungsgemeinschaft (DFG)  (S4  CRC 1119 CROSSING).

\section*{Acknowledgments}
The author is grateful to Prof. P. W. H. Pinkse, Prof. B.  \v{S}kori\'{c} and Prof. A. P. Mosk for their hospitality, and enlightening discussions. He is also grateful to Lukas Fladung and Yannick Deller, for their comments on the manuscript.


\bibliography{PUFComitOE}

\end{document}